\documentclass{article}
\usepackage[utf8]{inputenc}
\usepackage{float}
\usepackage{amsmath}
\usepackage{enumitem}
\usepackage[margin=1in]{geometry}
\usepackage{graphicx}
\usepackage[numbers]{natbib}
\usepackage{lineno}
\usepackage{svg}
\usepackage{subfigure}
\usepackage{upgreek}
\usepackage{hyperref}
\usepackage{multibib}
\newcites{SM}{SI References}

\title{Inferring Chemical Disequilibrium Biosignatures for Proterozoic Earth-Like Exoplanets}
\author{Amber V. Young,$^{1\ast}$ Tyler D. Robinson,$^{2}$ Joshua Krissansen-Totton,$^{3}$ Edward W. Schwieterman,$^{4}$ \\ Nicholas F. Wogan,$^{5}$ Michael J. Way,$^{6,7}$ Linda E. Sohl,$^{8,6}$ Giada N. Arney,$^{9}$ \\ Christopher T. Reinhard,$^{10}$ Michael R. Line,$^{11}$ David C. Catling,$^{3,5}$ James D. Windsor$^{1}$ \\
\normalsize{$^{1}$Department of Astronomy and Planetary Sciences, Northern Arizona University,}\\
\normalsize{Physical Sciences Building, 527 S Beaver St, Flagstaff, AZ 86011}\\
\normalsize{$^{2}$Lunar and Planetary Laboratory, University of Arizona, Tucson AZ 85721}\\
\normalsize{$^{3}$Earth and Space Sciences, University of Washington, Seattle WA 98195}\\
\normalsize{$^{4}$Department of Earth and Planetary Sciences, University of California Riverside, Riverside CA 92521}\\
\normalsize{$^{5}$Department of Astrobiology, University of Washington, Seattle WA 98195}\\
\normalsize{$^{6}$NASA Goddard Institute for Space Studies, New York NY 10025}\\
\normalsize{$^{7}$Department of Physics and Astronomy, Uppsala University, Uppsala, SE-75120, Sweden}\\
\normalsize{$^{8}$Center for Climate Systems Research, Columbia University, New York, New York, 10025}\\
\normalsize{$^{9}$NASA Goddard Space Flight Center, Greenbelt MD 20771}\\
\normalsize{$^{10}$Earth and Atmospheric Sciences, Georgia Tech, Atlanta GA 30332}\\
\normalsize{$^{11}$School of Earth and Space Exploration, Arizona State University, Tempe AZ 85281}\\
\normalsize{$^{12}$Center for Climate Systems Research, Columbia University, New York, New York, 10025}\\
\normalsize{$^\ast$To whom correspondence should be addressed: Amber\_Young86@nau.edu.}}

\begin{document}

\maketitle

\textbf{Chemical disequilibrium quantified via available free energy has previously been proposed as a potential biosignature. However, exoplanet biosignature remote sensing work has not yet investigated how observational uncertainties impact the ability to infer a life-generated available free energy. We pair an atmospheric retrieval tool to a thermodynamics model to assess the detectability of chemical disequilibrium signatures of Earth-like exoplanets, emphasizing the Proterozoic Eon where atmospheric abundances of oxygen-methane disequilibrium pairs may have been relatively high. Retrieval model studies applied across a range of gas abundances revealed that order-of-magnitude constraints on disequilibrium energy are achieved with simulated reflected-light observations at the high abundance scenario and signal-to-noise ratios (50) while weak constraints are found at moderate SNRs (20\,--\,30) for med\,--\,low abundance cases. Furthermore, the disequilibrium energy constraints are improved by modest thermal information encoded in water vapor opacities at optical and near-infrared wavelengths. These results highlight how remotely detecting chemical disequilibrium biosignatures can be a useful and metabolism- agnostic approach to biosignature detection. 
}

Exoplanet exploration science is making rapid progress toward the detection and characterization of potentially habitable worlds \citep{James_Webb_Reference}. Ongoing \citep{Greene_2016,JWST_CO2_Detect_Paper_2022} and near-future exoplanet strategies \citep{NAS_2021} will emphasize the search for atmospheric gases, including the chemical signatures of life (or biosignatures) \citep{Schwieterman_2018,Meadows_2018,Madhusudhan_2019}. Recently, the Decadal Survey on Astronomy and Astrophysics 2020 report \citep{NAS_2021} recommended space-based high contrast imaging of potentially life-bearing exoplanets as a leading priority for the coming decade. When attempting to infer if a distant world is inhabited, chemical disequilibrium is a potential indicator of life that has a long history of study in solar system planetary environments \citep{lovelock_physical_1965,Hitchcock_1967,Lovelock_1975}. A key example is the coexistence of O$_2$ and CH$_4$ in Earth's atmosphere where the strong biological production rates of CH$_4$ are able to maintain this gas at appreciable levels despite its relatively short chemical lifetime (roughly a decade) in an oxidizing atmosphere. 

A primary metric for quantifying chemical disequilibrium involves calculating the difference in chemical energy associated with an observed system and that system's theoretical equilibrium state. Recent work has explored the application of one such metric\,---\,the available Gibbs free energy\,---\,to solar system worlds and to Earth's planetary evolution \citep{krissansen-totton_detecting_2016,Krissansen-Totton_2018, Wogan_2020}. Although the available Gibbs free energy is a promising metric for interpreting chemical disequilibrium biosignatures, little is known about how observational uncertainties will impact our ability to constrain the available Gibbs free energy for Earth-like exoplanets. Where ``Earth-like" refers to an ocean bearing, Earth-sized world with surface pressures and temperatures similar to Earth and with an atmosphere dominated by N$_2$, H$_2$O, and CO$_2$ with trace amounts of CH$_4$ and varying levels of O$_2$.   

Exoplanet atmospheric characterization, including the search for biosignature gases, proceeds through retrieval analysis or atmospheric inference \citep[e.g.,][]{Madhusudhan_2009,Benneke_2012,Line_2013,Feng_2018,Barstow_2020}. In short, a retrieval framework enables the statistical exploration of atmospheric states that are consistent with a given set of spectral observations, be these real or simulated. While retrieval models do not directly constrain quantities like the available Gibbs free energy, pairing an exoplanet atmospheric retrieval model with a thermochemical tool\,---\,as detailed in Methods\,---\,enables inferences of both atmospheric chemical abundances and the associated disequilibrium state of the atmosphere. As described in Methods, simulated reflected-light observations of an Earth-like planet are created with uncertainties specified by the V-band (.55\,$\upmu$m) SNR. Applying inverse modeling techniques to these simulated observations for multiple randomized observational noise realizations then maps observational quality to expected constraints on the available Gibbs free energy.

Analyses presented here emphasize directly-imaged Proterozoic Earth analogs with reflectance spectral data spanning near-infrared/optical/ultra-violet wavelengths at resolving powers of 70/140/7 (motivated by Decadal Survey mission concept reports \citep{HabEx_Study_2018,LUVOIR_Study_2018}). This eon represents roughly half of Earth's history and is notable for its oxygenated atmosphere that also may have had enhanced atmospheric methane concentrations (as compared to modern Earth), thereby presenting an ideal time period for detecting O$_2$-CH$_4$ disequilibrium. Retrieval studies below explore a range of concentrations for key gases in Proterozoic Earth's atmosphere and were adopted from a span of Earth evolutionary scenarios summarized in a  review \citep{Robinson_Reinhard_2020}. High and low concentration scenarios here are identical to mid-Proterozoic extremes from this review, and an intermediate concentration case was generated by computing the logarithmic geometric mean of the high and low cases.  

\begin{figure}[H]
    \hspace{-1cm}
    \includegraphics[scale=0.50]{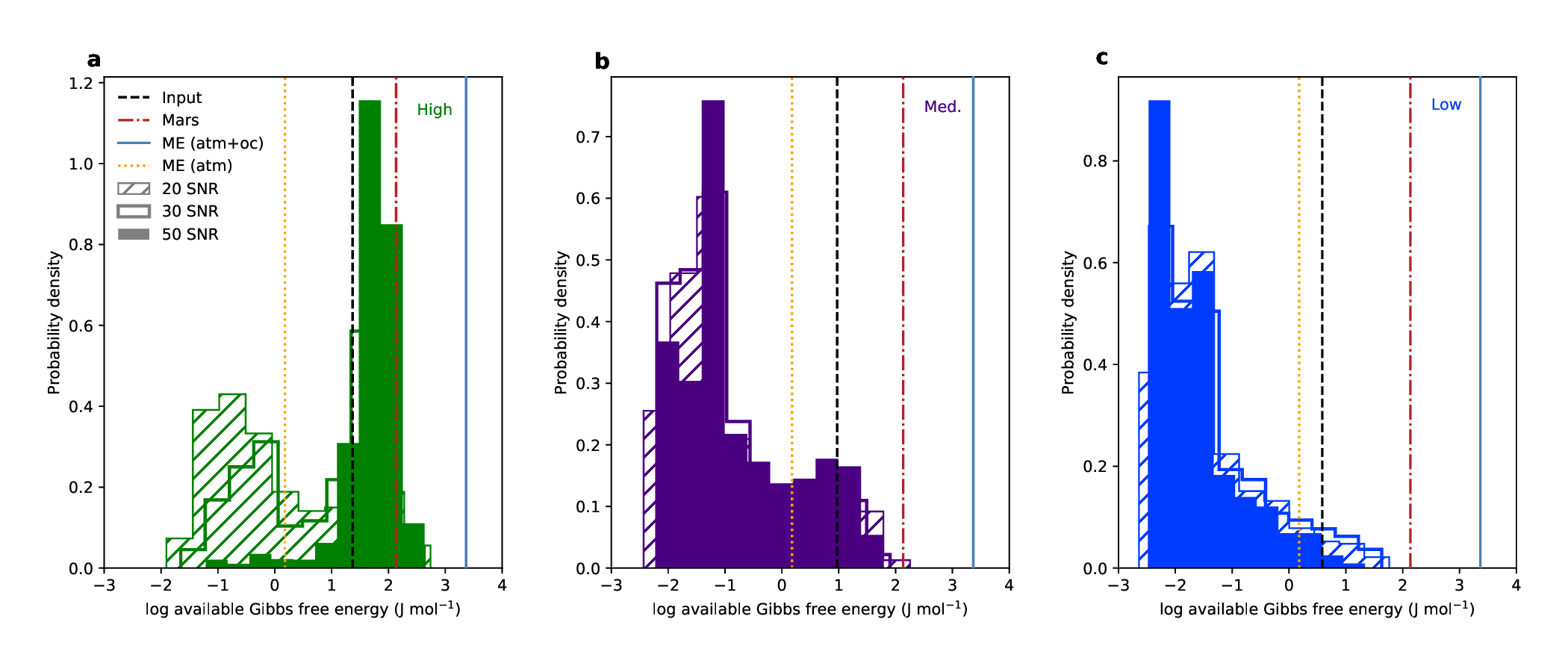}
    \caption{\textbf{Available Gibbs Free Energy Posterior Distributions Inferred from Simulated Reflected Light Observations for Different Proterozoic Earth Models.} \textbf{a}, The marginal posterior distribution of the log of the available Gibbs free energy for the high abundance case, derived from 20 (hatched), 30 (un-filled) and 50 (solid fill) SNR simulated reflected light observations. Vertical black (dashed), orange (dotted), red (dot-dashed), and blue (solid) lines in all three panels represent the input value and previously reported values for the available Gibbs free energy of modern Earth (atmosphere only case), Mars, and modern Earth (ME) (atmosphere-ocean case) respectively \citep{krissansen-totton_detecting_2016}. \textbf{b}, Same as \textbf{a} but for the medium abundance case. \textbf{c}, Same as \textbf{a} but for the low abundance case.}
    \label{Gibbs_Posteriors}
\end{figure}

Fig.~\ref{Gibbs_Posteriors} shows modeled constraints on the atmospheric available Gibbs free energy (in Joules per mole of atmosphere) that would be expected from observations of Proterozoic Earth analogs in reflected light. Each result is broken up into three atmospheric composition categories of ``high", ``medium", and ``low" biosignature gas abundances and simulated observations were conducted at several SNRs for each abundance category. Most of these reflected light cases present available Gibbs free energy posteriors that are consistently peaked at lower Gibbs free energy values but with statistically significant tails to higher values. In our simulations, the log available Gibbs free energy is found to be no larger than 1.30/1.13/1.21\,J\,mol$^{-1}$ at 95\% confidence (inferred from the marginal cumulative distributions) for SNRs of 20/30/50 for the medium abundance case and 0.47/0.68/0.03\,J\,mol$^{-1}$ for the low abundance case. By contrast, the uncertainty on the available Gibbs free energy for the high abundance case goes down to as low as an order of magnitude for the SNR 50 observational case (See Table \ref{1_sig_Table}). In the high abundance scenario (Fig. \ref{Gibbs_Posteriors}a), the posteriors derived from the simulated SNR 20 and 30 observations exhibit a dual peak in the distribution. Given the randomization of the simulated observational data points, retrievals at these SNRs could occasionally constrain the O$_2$ abundance. Thus, one peak corresponds to cases where O$_2$ was well-detected and the other peak corresponds to non-detections.

\begin{figure}[H]
    \centering
    \includegraphics[scale=0.6]{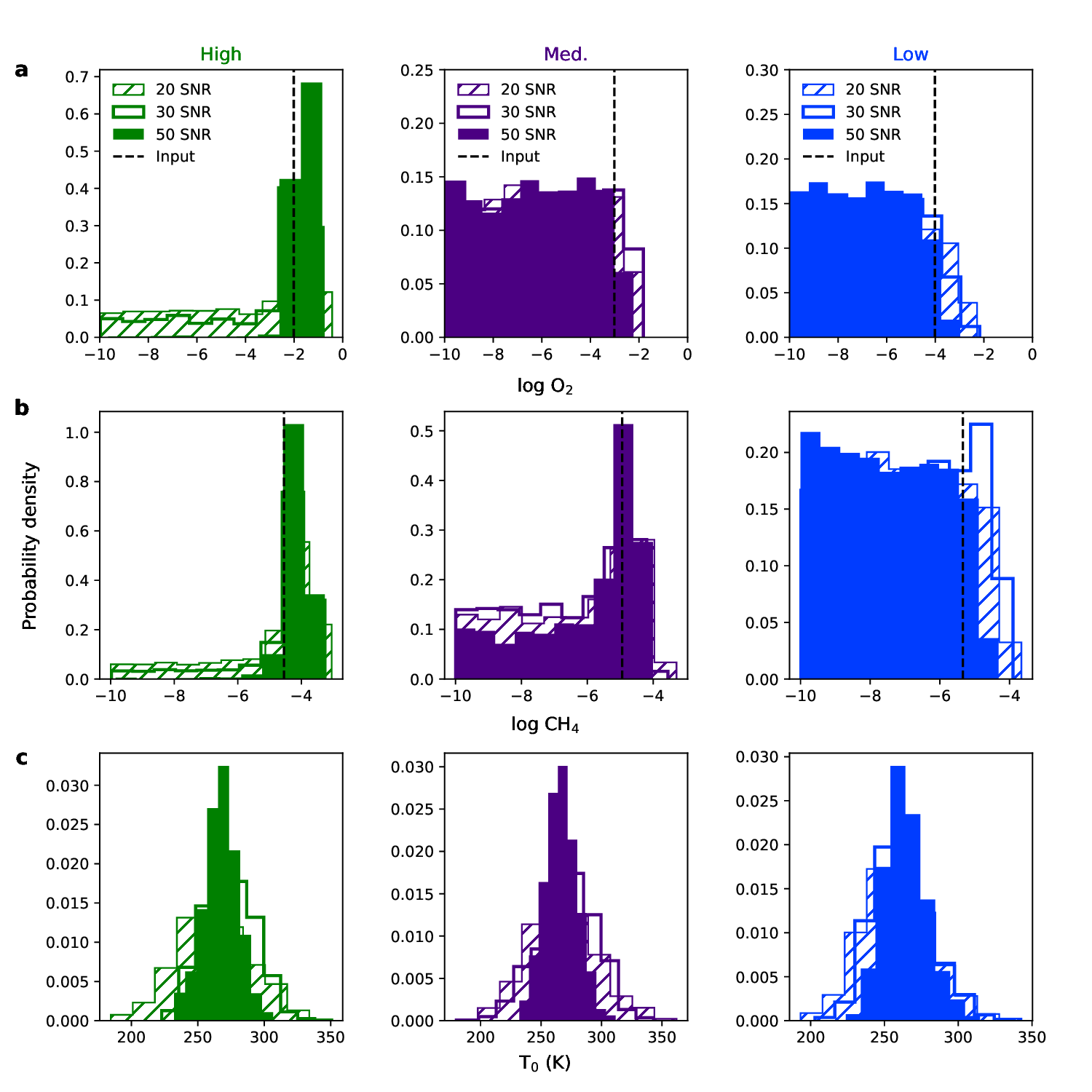}
    \caption{\textbf{Posterior Distributions for Key Retrieved Atmospheric Parameters for Different Noise Levels and Proterozoic Earth Models.} \textbf{a}, The marginal posterior probability distributions for the retrieved log abundance of O$_2$ at the high (green), medium (purple), and low (blue) abundance cases. Each distribution is inferred from simulated reflected light observations at SNRs of 20 (hatched), 30 (un-filled), and 50 (solid fill). A vertical black dashed line represents the input value for each parameter (the input is calculated via column integrated volume mixing ratio profiles for each gas phase species). \textbf{b}, Same as \textbf{a} except now showcasing methane constraints. \textbf{c}, Same as \textbf{a} except now showcasing atmospheric temperature constraints.}
    \label{R_Retrieval_Params}
\end{figure}

The constraints on the available Gibbs free energy are most-strongly dependent on the quality of the inferences for O$_2$ abundance, CH$_4$ abundance, and atmospheric temperature, which are shown in Fig.~\ref{R_Retrieval_Params}. This is consistent with thermodynamic theory, which has shown that the Gibbs free energy is strongly dependent on temperature and only weakly dependent on pressure \citep{engelthomas_thermodynamics_2019}. In the high abundance case at SNR 20, results show a large uncertainty on the inferred O$_2$ abundance. This introduced a significant uncertainty on the available Gibbs free energy for this particular case. However, higher observational SNRs of 30 and 50 at high abundance showed better constraints on O$_2$ which led to improved constraints on the available Gibbs free energy. These trends also held true for the CH$_4$ posteriors in the high abundance case. The retrieval analyses for the medium and low cases largely resulted in broad upper limit constraints for O$_2$ and CH$_4$ at each of the observational signal-to-noise scenarios tested here. Reasonable atmospheric temperature constraints were seen for all abundance cases and observing scenarios, which stemmed from adequate constraints on the shape of atmospheric water vapor bands across the spectral range for all modeled scenarios. Table~\ref{1_sig_Table} details the $16^{\rm th}$-, $50^{\rm th}$-, and $84^{\rm th}$-percentile values for the marginal O$_2$, CH$_4$, and temperature distributions (corresponding to the 1-sigma values for a Gaussian distribution).

\begin{figure}[H]
    \centering
    \includegraphics[scale=0.45]{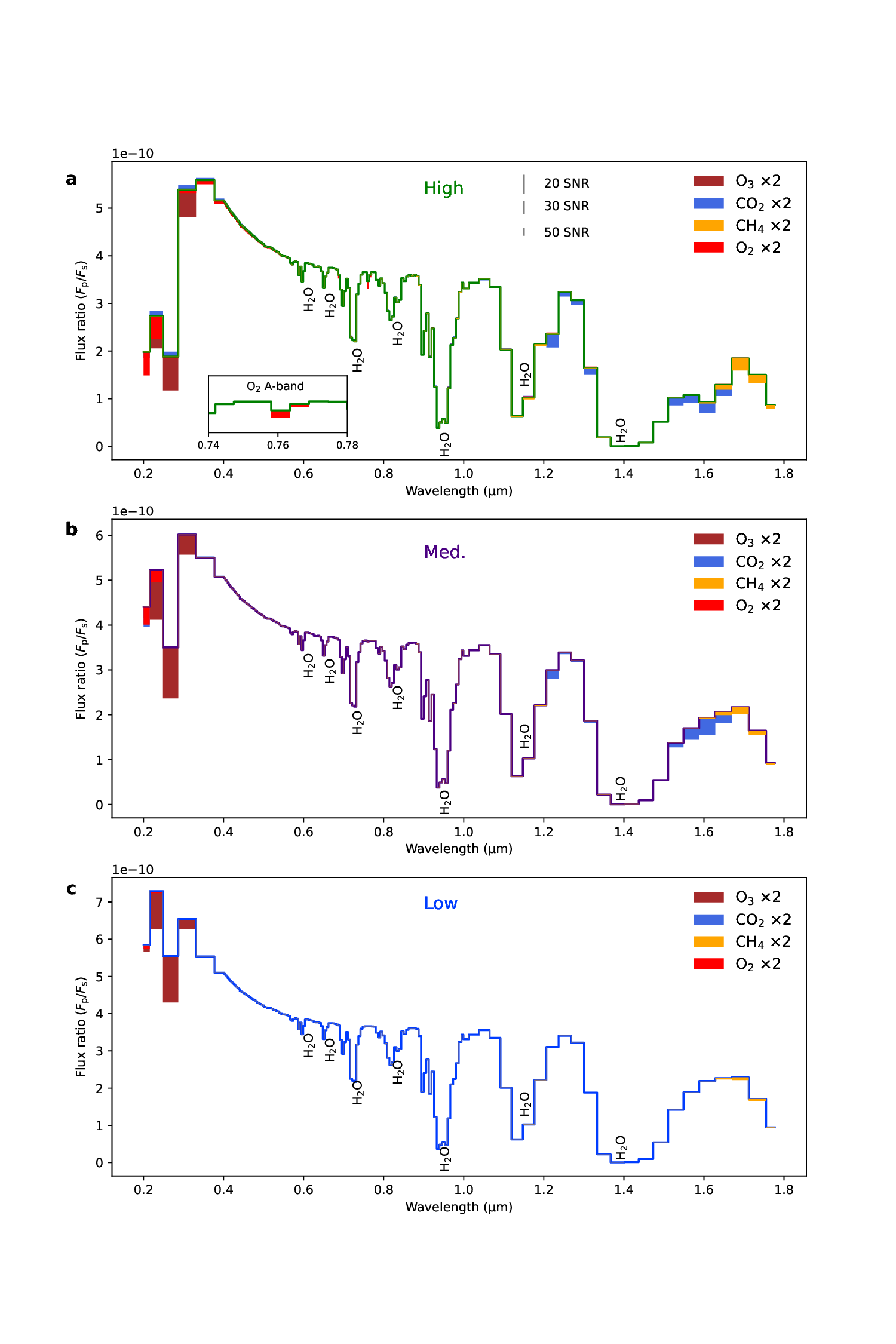}
    \caption{\textbf{Simulated Reflected Light Spectra for Proterozoic Earth Cases.} The x-axis indicates the wavelength in\,$\upmu$m and the y-axis represents the planet-to-star flux ratio. \textbf{a}, Simulated reflected light spectrum for the high (green) abundance case. Absorption features for O$_3$ (brown), CO$_2$ (blue), CH$_4$ (yellow), and O$_2$ (red) are shown and their input abundances are multiplied by a factor of two ($\times 2$). Water vapor absorption features are labeled with text as well. The grey error bar legend shows scaling with each noise instance (20, 30, and 50 SNR). Also included is an inset figure highlighting the O$_2$ A-band feature located at 0.76\,$\upmu$m. \textbf{b}, The simulated reflected light spectrum for the medium (purple) abundance case. Each input abundance is multiplied by a factor of two ($\times 2$) to show its effect. \textbf{c}, The simulated reflected light spectrum for the low (blue) abundance case. The denoted species absorption features are shown and their input abundances are multiplied by a factor of two ($\times 2$).}
    \label{RT_Spectra}
\end{figure}

Most fundamentally, results for the high abundance scenario demonstrated that strong detections of O$_2$ and CH$_4$ absorption features lead to tight constraints on the resulting available Gibbs free energy, thereby enabling an inference of the extent of chemical disequilibrium in the atmosphere of an Earth-like exoplanet. Fig.~\ref{RT_Spectra} highlights spectral features of several species including O$_2$, CH$_4$, O$_3$, CO$_2$, and H$_2$O across the range of modeled Proterozoic Earth scenarios. In the near-infrared/optical/ultra-violet spectral range explored in this work, the strongest O$_2$ feature is the oxygen A-band at 0.762\,$\upmu$m. There are several CH$_4$ features within the 1.6 - 1.8\,$\upmu$m wavelength range, indicated 
in orange. Each color coded absorption feature for O$_3$, CO$_2$, CH$_4$, and O$_2$ is accentuated by a factor of two relative to the original input abundance in order to highlight the precision needed to observe each species. 

Constraining the available Gibbs free energy is a promising characterization strategy that synergizes well with established techniques for biosignature gas detection. In practice, it is possible to infer an upper limit on the available Gibbs free energy and, for more optimistic cases, proper constraints can be obtained on the free energy for Proterozoic Earth-like planets. Particularly for high abundance cases, it is possible to place constraints on the available Gibbs free energy to within an order of magnitude with SNR 50 observations. This could be feasible with a future exoplanet direct imaging mission, but future work is necessary to understand any systematic barriers to achieving this level of signal-to-noise. It is also worth noting that the detection of the O$_2$-CH$_4$ disequilibrium is highly sensitive to the near-infrared spectral features that drive the quality of the CH$_4$ abundance constraints and is also, in part, enabled by atmospheric temperature constraints from thermal effects in molecular band shapes (especially in the near-infrared). Such spectral features may not be observable for all targets as the inner working angle for high contrast imaging systems, especially coronagraphs, expands linearly with wavelength, thus further indicating the importance of a small inner working angle for a future mission. Features shortward of 1.8\,$\upmu$m observed with a 6\,m telescope would be accessible for targets within $\sim8$ pc. For a 6\,m-class space telescope with noise properties modeled on the LUVOIR-B concept \citep{LUVOIR_Study_2018,Robinson_2016}, the high-SNR cases explored here could be achieved for an Earth-like target around a solar host at distances of 5—7\,pc with an investment of 2—4 weeks of observing time (in line, for example, with planned expenditures for high-value targets in the HabEx concept study \citep{HabEx_Study_2018}). From a recently proposed target star list for HWO \citep{HWO_Star_List}, about 26 stars are within that 7\,pc limit. Two important observation-related caveats are that the retrievals assumed the planetary orbit to be well-constrained at that planet-to-star flux ratio, and could be treated as time-independent over the course of an observation. As an observing strategy, performing a baseline analysis at lower SNRs may help us identify potentially exciting targets for more detailed follow-up observations and provide upper limit constraints on potential chemical disequilibrium signals for a subset of targets.

From an observational perspective, characterizing the CH$_4$ and O$_2$ abundances are essential to inferring the atmospheric chemical disequilibrium signal of Earth analogs over most of its evolutionary history. While results presented here emphasize the detectability of O$_2$ and CH$_4$ from space with reflected-light observations at visible and near-infrared wavelengths, similarly effective observations could be done from the ground and in other spectral ranges. For example, Extremely Large Telescopes could attempt O$_2$ detections for Earth-like planets orbiting nearby M dwarfs \citep{Kawahara_2012,Rodler_2014,Hardegree_2023,Currie_2023}. Ground based observations would compliment ongoing efforts from the \textit{James Webb Space Telescope} where constraining Earth-like abundances of O$_2$ is a known challenge \citep{Wunderlich_2019}. Analogous transit scenarios relevant to the \textit{James Webb Space Telescope} NIRSpec instrument are explored in the Supplementary materials (see Supplementary Figure 1 and Supplementary Figure 2) and show the available Gibbs free energy is difficult to constrain (most likely due to non-detections of O$_2$ from the simulated observations). Conversely, molecular oxygen may be detectable in the mid-infrared with transit spectroscopy via observing O$_2$ collision-induced absorption features near 6.4\,$\upmu$m with the \textit{JWST} MIRI instrument  \citep{Fauchez_etal_2020}. While O$_2$ does not produce strong features in the emission spectrum of an Earth-like exoplanet, if O$_2$ abundances could be inferred from O$_3$ abundances \citep{Kozakis_2022} then detections of O$_3$ in the mid-infrared at 9.7\,$\upmu$m with low/moderate resolution spectroscopy could provide the requisite constraints on O$_2$. The LIFE (Large Interferometer for Exoplanets) mission concept, for example, could constrain O$_3$ abundances at SNRs $>$ 10 in the mid-infrared \citep{Konrad_2022_LIFE3}. Ozone also has absorption features in the ultra-violet (e.g., the Hartley-Huggins band at 0.25\,$\upmu$m and the subtler Chappuis bands from 0.5 to 0.7\,$\upmu$m), which could be detectable with HWO using low/moderate resolution reflected light observations as well (See Supplementary Figure 3c). CH$_4$ has key absorption bands throughout the optical, near-infrared, and mid-infrared, so could be detected via reflectance, transmission, and/or emission spectroscopy and would require resolving powers of roughly 30—40, depending on abundance \citep[e.g.,][]{Robinson_Salvador_2023,Wunderlich_2019,Gialluca_2021,Currie_2023,Quanz_2022}. 

The Proterozoic Eon is a potentially ideal Earth-like context for constraining the atmospheric O$_2$-CH$_4$ chemical disequilibrium gas pair for an Earth-like planet around a G-type star due to a likely higher abundance of CH$_4$, a rise in O$_2$ relative to the Archean Earth, and the longevity of this signal over a 2\,Gyr time period. An Archean Earth-like atmosphere may have a modest atmospheric chemical disequilibrium signature driven by the CH$_4$-CO$_2$ gas pair \citep{Krissansen-Totton_2018}. However, abiotic sources for CH$_4$ production would need to be explored \citep{Krissansen-Totton_2018}. Modern Earth has substantially less atmospheric CH$_4$, making detection of the O$_2$-CH$_4$ atmospheric signature challenging, although the photochemistry for a modern Earth-like planet around an M- or K-dwarf may generate substantially more CH$_4$ in atmospheres with modern levels of O$_2$ \citep[e.g.,][]{Segura_2005,Arney_2019}. Thus, the concept of remotely-detectable disequilibrium energy biosignatures could apply to Earth-like worlds around a very wide range of stellar host types. For any exoplanet, the stellar photochemical context will be vital to consider when evaluating potential biosignatures. Nevertheless, our results offer a window into the characterization of an Earth-Sun twin as an analog for similar exoplanets.

The thermodynamic systems modeled here only represent the chemical disequilibrium in the atmosphere. Oceans can provide an additional source for chemical disequilibrium and, in fact, the maintenance of N$_2$ and O$_2$ in the presence of liquid water (for Proterozoic and modern Earth) and the maintenance of CO$_2$, N$_2$, and CH$_4$ in the presence of liquid water (for Archean Earth) are major contributors to disequilibrium energy over time \citep{Krissansen-Totton_2018}. There is potential to remotely detect exoplanetary oceans in reflected light \citep{Cowan_2009,Robinson_2010_glint,Zugger_2010,Lustig-Yaeger_2018}. However, inference of available energy requires constraints on the planet's ocean volume, which may be quite challenging to remotely constrain. Thus, atmospheric disequilibrium constraints are likely to be conservative, at least for ocean-bearing worlds. In general, the inability to easily constrain ocean volume (and/or its chemical state) or the inability of retrieval to fully detect all gaseous species makes the atmospheric available Gibbs free energy constraints likely to be conservative. 
 
Discriminating false positives, where an abiotically driven signal could mimic a true biosignature, is especially crucial for interpreting chemical disequilibrium signatures considering false positives for O$_2$ have been heavily studied \citep{Harman_2018,Luger_Barnes_2015,Domagal_Goldman_2014,Wordsworth_2014,Krissansen_Totton_2021}. The holistic interpretation of a given chemical disequilibrium signal will not only require quantifying the extent of the signal, but also inferring the chemical species driving the signature and their production/loss mechanisms \citep{Meadows_2018}. A previous study by \citet{Wogan_2020} outlined the potential abiotic mechanisms for generating free energy, and even cases where a lack of free energy could lead to a false negative scenario. 

A known environment with an excess amount of abiotically produced free energy is Mars and in fact estimates of chemical disequilibrium for the Proterozoic Earth are \textit{lower} than the modern Mars case (24\,J\,mol$^{-1}$ versus 136\,J\,mol$^{-1}$, respectively). However, the available Gibbs free energy produced on Mars is driven by the O$_2$-CO gas pair (abiotically generated by CO$_2$ photolysis), while abiotic sources for free energy in the Proterozoic Earth atmosphere are negligible (Supplementary Figure 4). While the direct imaging characterization of a true Mars analog exoplanet would be challenging due to the tenuous nature of the atmosphere, a larger orbital distance than Earth-like targets, and overall smaller size, Mars-like worlds with substantial available Gibbs free energy due to photochemistry may represent an important category of false-positives. Mars exhibits very cold, dry atmospheric conditions with a majority of its atmosphere comprised of CO$_2$, which can be readily photolyzed. Given this context, one would need to constrain the photochemical environment of a Mars analog in order to distinguish it from a Proterozoic Earth-like case. This would require atmospheric constraints (e.g., species abundances, planetary surface pressure, and atmospheric temperature) obtained alongside characterizing the host stellar spectrum (especially at ultraviolet wavelengths). The abundances of CO$_2$, CO, and abiotic O$_2$ would be especially important to characterize. CO$_2$ and CO have absorption features in the near-infrared around 1.6\,$\upmu$m and CO$_2$ has additional, weaker features at shorter near-infrared wavelengths. CO absorption in this wavelength range is particularly weak and would likely require high CO partial pressures to be detectable, thus implying that it might be difficult to detect abiotic disequilibrium energy for such worlds. 

In any  given search, a planet's retrieved chemical disequilibrium must be considered alongside other contextual information to establish biogenicity. For example, a disequilibrium that requires gas fluxes incompatible with abiotic explanations is more likely to be due to life. Additionally, planets with ``edible" disequilibria (i.e., easily surmountable kinetic barriers) that might be expected to be readily consumed as a metabolic fuel by a resident biosphere may serve as ``anti-biosignatures" \citep{Wogan_2020}. In general, detection of atmospheric chemical disequilibrium would require that the associated chemical species be well-mixed in the atmosphere while more-localized, non-global signatures would be much more difficult to constrain. 

In summary, searching for chemical disequilibrium biosignatures will be a promising endeavor for exoplanet characterization efforts and future missions should consider this approach when developing their observational strategies. For Earth-like analogs, access to O$_2$ and CH$_4$ spectral features is key and will hinge on sufficient resolving power and broad spectral coverage to retrieve relevant gas features. Maximizing the potential to place tight constraints on these chemical species will likely require high SNR observations at wavelengths spanning into the near-infrared, implying this technique may be most relevant to the best exoplanetary candidates. Alongside understanding planets and stars as systems \citep{Meadows_2018}, chemical disequilibrium biosignatures will be a powerful tool for future observations. 

\section*{Methods}
This study incorporated an exoplanet atmospheric retrieval model that was coupled to a thermodynamics Gibbs free energy tool to explore how observational quality influences our ability to interpret and quantify chemical disequilibrium signals from simulated reflected light observations. The photochemical model, \texttt{Atmos}, was used to explore a range of atmospheric compositions spanning high, medium, and low biosignature gas abundance and were  then used to generate simulated spectra according to each abundance case. Thereafter a spectral retrieval model, \texttt{rfast}, was used to map out the posterior distributions of relevant atmospheric and planetary parameters consistent with a given simulated observation. The \texttt{rfast} and the Gibbs free energy tools were then coupled via randomly sampling the atmospheric state posterior distribution for relevant parameters and passing those randomized instances as inputs to the Gibbs free energy tool. Repeating this process thousands of times generates a posterior distribution for the available Gibbs free energy that is consistent with the simulated observation. The resulting posterior distribution allowed us asses how observational uncertainty influences our ability to constrain and interpret chemical disequilibrium biosignatures. 

\subsection*{The Retrieval Model}
The \texttt{rfast} model \citep{Robinson_Salvador_2023} was adopted for the atmospheric retrievals. It incorporates a radiative transfer forward model, an instrument noise model, and a retrieval tool to enable rapid investigations of exoplanet atmospheric remote sensing scenarios. The radiative transfer forward model is capable of simulating (1) both 1-D and 3-D views of an exoplanet in reflected light, (2) emission spectra, and (3) transit spectra, and takes as input the atmospheric chemical and thermal state (including profiles of cloud properties).The retrieval package also uses a Bayesian sampling package (\texttt{emcee}) to call the aforementioned radiative transfer model while mapping out the posterior distribution for the atmospheric parameters used to fit a noisy observation \citep{Foreman_Mackey_2013}.

Our retrieval simulations are done in 1D using the diffuse radiative transfer treatment for modeling directly imaged views of the planet, and reproduces what one might expect for a quadrature (or gibbous) phase observation of an Earth-like exoplanet. The \texttt{rfast} model performs the retrieval inference assuming an isothermal atmosphere, constant volume mixing ratio profiles for input chemical species, and clouds are taken to be an equal blend of liquid water and ice. Additionally, clouds in simulated observations are assumed to cover 50\% of the disk and have a vertical extent described by the cloud top height and cloud pressure extent in \citet{Feng_2018}. Our noise estimates are constant with wavelength and set by the noise specified at V-band in order to remain consistent with decadal studies and prior exo-Earth retrieval analyses \citep{HabEx_Study_2018,LUVOIR_Study_2018,Feng_2018}.  

\subsection*{Gibbs Free Energy Model}
The thermodynamics Gibbs free energy model from \citet{krissansen-totton_detecting_2016} and \citet{Krissansen-Totton_2018} was adopted to calculate chemical disequilibrium biosignatures. The Gibbs free energy is a thermodynamic state function that describes the maximum amount of work a chemical process can produce at constant pressure/temperature \citep{engelthomas_thermodynamics_2019}: 

\begin{equation}
    G = \sum_{i}^{N} \left(\frac{\partial G}{\partial n_i}\right)n_i
\end{equation}
where $\rm n_i$ is the moles of species `$i$', $\partial G/\partial n_i$ is the change in Gibbs free energy with respect to the moles of a given species, and the sum is over the total number of species in the system ($N$). The overall Gibbs free energy of a given state can be rewritten in terms of thermodynamic activity and the standard Gibbs free energy of formation for a given species: 

\begin{equation}
    \Delta G_{(T,P)} = \sum_{i}^{N} \left(\Delta_fG_{i(T,P_r)}^o + RT\ln{\left(\frac{P \rm n_i}{n_T}\gamma_{fi}\right)} \right)n_i
\end{equation}
where, $\Delta_fG_{i(T,P_r)}^o$ is the Gibbs free energy of formation for a given species (determined at a given global mean surface temperature, $T$, and reference pressure, $P_r$), $n_T$ is the total number of moles, and $\gamma_{fi}$ is the activity coefficient. The utility of Gibbs free energy is that it is minimized at thermodynamic equilibrium, which allows us to model the theoretical equilibrium state at a given temperature and pressure without explicitly considering individual chemical reactions.

Each observational scenario that was simulated considers the target as a closed, well-mixed system at constant global surface pressure and constant characteristic global temperature \citep{krissansen-totton_detecting_2016}. The Gibbs free energy model first computes the Gibbs free energy of the system given an input atmospheric state (e.g., usually an instance of gas mixing ratios, atmospheric temperature, and surface pressure from the atmospheric retrieval model) and subsequently solves for the equilibrium species mixing ratios, where the Gibbs free energy is minimized and atoms are conserved. An interior points method, implemented using Matlab's fmincon function, was used to minimize and solve for the equilibrium state (See Supplementary Figure 4 which details the initial and equilibrium abundances for all the species). The difference between the Gibbs energy of the input atmospheric state and the equilibrium state quantifies the ``available Gibbs free energy'': 

\begin{equation}
   \Phi \equiv G_{(T,P)}(n_{\rm initial}) - G_{(T,P)}(n_{\rm eq}) 
\end{equation}
The available Gibbs free energy ($\Phi$) is henceforth used to quantify chemical disequilibrium, and is measured in Joules of available free energy per mole of atmosphere \citep{krissansen-totton_detecting_2016,Krissansen-Totton_2018}. A large available Gibbs free energy indicates strong planetary chemical disequilibrium (e.g., the modern Earth atmosphere-ocean system at 2,326\,J\,mol$^{-1}$). Conversely, low available Gibbs free energy indicates weak chemical disequilibrium. Most planetary bodies in our solar system like Jupiter, Venus, and Uranus, all have well below 1\,J\,mol$^{-1}$ of available free energy \citep{krissansen-totton_detecting_2016}.

\subsection*{Thermodynamics Calculations}
Supplementary Figure 4 shows an equilibrium calculation for the high Proterozoic abundance case presented in the main text. O$_2$, N$_2$, H$_2$O, CO$_2$, NH$_3$, CH$_4$, H$_2$, N$_2$O, and O$_3$ are all included in the calculation. The total available Gibbs free energy for this scenario is 24.24\,J\,mol$^{-1}$. The blue bars show the observed mixing ratios of each species, the red bars represent the equilibrium mixing ratios of each species, and the yellow bars show the change in mixing ratio between the equilibrium and observed abundances. O$_2$ and CH$_4$ have significantly large observed mixing ratios and also have large differences between their observed and equilibrium abundances. This indicates that these two species are the main drivers of chemical disequilibrium in the atmosphere. NH$_3$, H$_2$, N$_2$O, and O$_3$ all contribute to to the overall available free energy, however their low relative abundances (in comparison to O$_2$ and CH$_4$) make it such that the contribution of free energy from these species is negligible. Therefore, NH$_3$, H$_2$, and N$_2$O were excluded from the retrieval analysis and held at their equilibrium abundances when computing the available Gibbs free energy posterior distribution. O$_3$ was kept in the analysis since it is a photochemical byproduct of O$_2$ and thus useful to constrain alongside O$_2$.

\subsection*{Proterozoic Earth Atmospheric Modeling}
Self-consistent atmospheric models were generated using the photochemical model component of the \texttt{Atmos} tool \citep{arney_pale_2016,Arney_pale_2017}. \texttt{Atmos} is a coupled photochemical-climate model that uses planetary inputs (e.g., chemical species mixing ratios, associated chemical reactions, gravity, surface pressure, surface temperature, and stellar spectrum) to calculate the steady-state profiles of chemical species present in the atmosphere. In the overall analysis, the default solar spectrum was used to model Earth-like cases relevant to reflected light observations \citep{Thullier_etal_2004}. In the model, this solar spectrum was then adjusted with an input parameter (TIMEGA) to scale the solar flux to its appropriate insolation 1.3 Gyrs ago. The TRAPPIST-1 spectrum \citep{Peacock_2019} was used for the TRAPPIST-1 simulations that were mentioned in the text. All  the atmospheric cases modeled in this study assumed a total fixed surface pressure of 1\,bar. A suite of cases spanning low to high concentrations in O$_2$, CH$_4$, and CO$_2$ (input values outlined in Table \ref{1_sig_Table}) were explored to capture broad uncertainty in the abundance of certain atmospheric species during the Proterozoic eon and referenced to the``model low'' and ``model high" values from Table~1 in \citet{Robinson_Reinhard_2020}. The med. abundance case was computed taking the logarithmic geometric average of the high and low values. These generated profiles were used to produce realistic atmospheric spectra with the \texttt{rfast} forward model that were then retrieved on. Note that the changes in CO$_2$ abundance did not have a significant impact on the available Gibbs free energy uncertainty.

\subsection*{Simulated \texttt{rfast} Observations}
Reflected light observations in this study were modeled after direct imaging Decadal Survey mission concepts with ultra-violet (0.2--0.4\,$\upmu$m), optical (0.4--1.0\,$\upmu$m), and near-infrared (1.0--1.8\,$\upmu$m) bandpasses at resolving powers of 7, 140, and 70, respectively \citep{LUVOIR_Study_2018,HabEx_Study_2018}. Each instance of a simulated noisy spectrum was produced with randomized error bars and with the prescribed SNR taken to apply at 0.55\,$\upmu$m (consistent with earlier exo-Earth studies). 

\subsection*{Reflected Light Retrievals}
Outlined in Supplementary Figure 3 are the marginal posterior distributions for the mixing ratios of H$_2$O, Supplementary Figure 3a, CO$_2$, Supplementary Figure 3b, and O$_3$, Supplementary Figure 3c, along with the retrieved atmospheric pressure, Supplementary Figure 3d, for the high (green) med., (purple), and low (blue) abundance scenarios. These retrieved parameters were included in the random sampling used to compute the available Gibbs free energy posteriors (Fig.~\ref{Gibbs_Posteriors}), but did not have a significant influence on the uncertainty. However, constraining parameters like H$_2$O, CO$_2$, and surface pressure are essential for inferring the climate and habitability of exoplanets. Additionally, O$_3$ can be used as a proxy for loosely approximating the O$_2$ abundance for less oxygenated atmospheres since O$_3$ can remain detectable even at low O$_2$ concentrations \citep{Meadows_2017,Meadows_2018,Schwieterman_2018}; with the caveat that other factors (e.g., the stellar host type and stellar UV flux) are well characterized \citep{Kozakis_2022}. A comprehensive list of the planetary parameters included in the retrieval analysis are shown in Supplementary Table 1.

\subsection*{Transit Retrievals \& Available Gibbs Free Energy Inference}
The successful launch of the \textit{James Webb Space Telescope} (\textit{JWST}) and the prospects for characterizing Earth-like planets in the habitable zone of M dwarf stars motivated attempts to constrain the available Gibbs free energy of a Proterozoic Earth-like planet orbiting the M dwarf TRAPPIST-1. For all three atmospheric cases, and simulated observations with the NIRSpec instrument, results (Supplementary Figure 1 and Supplementary Figure 2) indicate that it is extremely challenging requiring less than 5\,ppm noise. 

In Supplementary Figure 1, the marginal posterior distributions for the O$_2$ abundance (Supplementary Figure 1a), CH$_4$ abundance (Supplementary Figure 1b), and atmospheric temperature (Supplementary Figure 1c) are outlined at the high, med, and low abundance cases for a Proterozoic Earth-like planet orbiting an M dwarf and inferred from simulated transit observations with the \textit{JWST} NIRSpec instrument. Particularly for O$_2$, these results show it is very difficult to constrain the atmospheric abundance of O$_2$ at each of the observational noise levels (5\,ppm and 10\,ppm), either of which would be challenging for \textit{JWST} to achieve for current best-case targets. This outcome was to be expected given that detecting biogenic O$_2$ abundances with \textit{JWST} is a known challenge \citep{Fauchez_etal_2020,Krissansen_Totton_TRAPPIST_2018,Lustig_Yeager_2019,Wunderlich_2019}. The lack of constraints on the O$_2$ abundance also make inferring the chemical disequilibrium energy of a Proterozoic Earth-like exoplanet orbiting a late-type star too challenging for \textit{JWST}. In Supplementary Figure 2, the available Gibbs free energy posteriors inferred from these observations are shown for the high (Supplementary Figure 2a), med. (Supplementary Figure 2b), and low (Supplementary Figure 2c) abundance cases. The posterior distributions for all abundance cases and noise levels demonstrate that a noise floor of $< 5$\,ppm would be required to obtain tight constraints on the available Gibbs free energy for these Proterozoic Earth-like scenarios. This $< 5$ \,ppm noise estimate is smaller than some predicted estimates (on the order of tens of ppm) for the noise floor  \citep{Greene_2016,Rustamkulov_2022}. It is therefore unlikely that \textit{JWST} could constrain the available Gibbs free energy for a Proterozoic Earth-like planet.

\section*{Data Availability}
All the source data that corresponds to the results of this work (e.g., spectral data, available Gibbs free energy distributions, and retrieval parameter distributions) were published to Zenodo \citep{Source_Data_2023} and available to download at \url{https://zenodo.org/record/8335447}. The initial release of the \texttt{rfast} spectral retrieval model can be accessed via \url{https://zenodo.org/record/7327817}. The \texttt{Atmos} model is publicly available and can be accessed at \url{https://github.com/VirtualPlanetaryLaboratory/atmos}. The Matlab version of the thermodynamics model from \citet{krissansen-totton_detecting_2016} and \citet{Krissansen-Totton_2018} are available from co-author Joshua Krissansen-Totton's personal website: \url{http://www.krisstott.com/code.html}.

\section*{Code Availability}
\texttt{rfast}: \url{https://zenodo.org/record/7327817}, \texttt{Atmos}: \url{https://github.com/VirtualPlanetaryLaboratory/atmos}, thermodynamics model: \url{http://www.krisstott.com/code.html}

\section*{Acknowledgements}
All authors would like to acknowledge support from the NASA Exobiology Program, and NASA's Nexus for Exoplanet System Science Virtual Planetary Laboratory. We would also like to acknowledge support from the NASA's Exoplanets Research Program, Habitable Worlds Program, and the NASA Interdisciplinary Consortia for Astrobiology Research (ICAR) Program via the Alternative Earths team. We acknowledge support from the Cottrell Scholar Program administered by the Research Corporation for Science Advancement. Finally, we acknowledge support from the GSFC Sellers Exoplanet Environments Collaboration (SEEC) and ROCKE-3D: The evolution of solar system worlds through time, both of which are funded by the NASA Planetary Science Divisions Internal Scientist Funding Model. This research was supported by Grant No.~80NSSC18K0349 (A.V.Y., T.D.R., J.K.T., E.W.S., M.J.W., L.E.S., G.N.A., C.T.R., M.R.L., D.C.C., and J.D.W.), No.~80NSSC18K0829 (A.V.Y., T.D.R., E.W.S., D.C.C., and N.F.W.), No.~80NSSC18K0349, No.~80NSSC20K0226 (T.D.R.), and No.~80NSSC21K059 (E.W.S. and C.T.R.).          

\section*{Author Contributions}
AVY conducted all the atmospheric retrievals outlined in this work, performed code development to couple the retrieval results to the Gibbs free energy model, and computed the available Gibbs free energy posteriors for all the simulated cases. TDR performed code adaptations and development of \texttt{rfast} in correspondence with the retrieval analyses performed in this work and is the principal investigator of the project. JKT distributed the most up to date version of thermodynamics model used in this work and contributed to necessary code development and adaptation for the thermodynamic simulations. ES provided \texttt{Atmos} generated atmospheric compositions for each of the abundance scenarios outlined in this work. All authors contributed to both project development and the writing and/or editing of the paper and approved the scientific content.

\section*{Competing Interests}
The authors declare no competing interests. 

\renewcommand{\arraystretch}{1.4}
\begin{table}[H]
    \centering
    \begin{tabular}{c c c c c c}
    \hline
    \hline
    &  &  \textbf{Input} & SNR 20 & SNR 30 & SNR 50  \\
    \hline 
    &  High & -2.01 & $-2.92_{-4.76}^{+1.45}$ & $-2.07_{-4.54}^{+0.49}$ & $-1.66_{-0.23}^{+0.24}$  \\
    log O$_2$ &  Med.  & -3.01 & $-6.08_{-2.67}^{+2.65}$ & $-6.02_{-2.71}^{+2.68}$ & $-6.29_{-2.52}^{+2.54}$ \\ 
    &  Low  & -4.01 & $-6.60_{-2.31
    }^{+2.37}$ & $-6.73_{-2.23}^{+2.25}$ & $-7.00_{-2.05}^{+2.05}$  \\
    \hline
    &  High  & -4.54 & $-4.36_{-3.04}^{+0.62}$ & $-4.20_{-1.28}^{+0.35}$ & $-4.11_{-0.32}^{+0.27}$  \\
    log CH$_4$ &  Med.  & -4.94 & $-6.07_{-2.66}^{+1.52}$ & $-6.29_{-2.52}^{+1.57}$ & $-5.31_{-2.94}^{+0.71}$  \\ 
    &  Low  & -5.53 & $-7.19_{-1.90}^{+1.92}$ & $-6.87_{-2.12}^{+1.92}$ & $-7.44_{-1.75}^{+1.76}$  \\
    \hline
    &  High  & -1.16 & $-0.65_{-0.39}^{+0.36}$ & $-0.79_{-0.26}^{+0.24}$ & $-0.75_{-0.20}^{+0.22}$  \\
    log CO$_2$ &  Med.  & -2.23 & $-2.28_{-3.27}^{+0.48}$ & $-2.14_{-0.34}^{+0.27}$ & $-2.19_{-0.27}^{+0.22}$  \\ 
    &  Low  & -3.31 & $-6.36_{-2.48}^{+2.54}$ & $-6.40_{-2.46}^{+2.44}$ & $-6.75_{-2.21}^{+2.25}$  \\
    \hline
    &  High  & 288. & $259.91_{-26.39}^{+32.18}$ & $272.23_{-20.62}^{+21.41}$ & $266.86_{-12.24}^{+12.44}$  \\
    T$_0$ (K) &  Med. & 288. & $267.84_{-27.70}^{+30.15}$ & $273.84_{-28.01}^{+22.47}$ & $266.13_{-12.44}^{+14.21}$  \\ 
    &  Low  & 288. & $255.31_{-19.79}^{+25.06}$ & $258.34_{-17.31}^{+23.07}$ & $263.09_{-13.10}^{+15.18}$  \\
    \hline
    &  High  & 1.3687 & $0.17_{-0.95}^{+1.69}$ & $1.28_{-1.72}^{+0.59}$ & $1.76_{-0.32}^{+0.28}$  \\
    log Avail. Gibbs (log \,J\,mol$^{-1}$) &  Med.  & 0.9739 & $-1.24_{-0.65}^{+1.59}$ & $-1.23_{-0.60}^{+1.54}$ & $-1.16_{-0.62}^{+1.74}$  \\ 
    &  Low  & 0.5853 & $-1.65_{-0.55}^{+1.11}$ & $-1.63_{-0.60}^{+1.18}$ & $-1.74_{-0.57}^{+0.81}$  \\
    \hline
    \hline 
   
    \end{tabular}
    \caption{Summary of the parameters, input values, and 1$\sigma$ confidence intervals (taken from the $16^{\rm th}$, $50^{\rm th}$, and $84^{\rm th}$ percentile values) for the high, med., and low atmospheric abundance scenarios that were modeled.}
    \label{1_sig_Table}
\end{table}


\end{document}